\documentclass[trackchanges]{aastex7}

\newcommand{\taylorcordes}{\texttt{Taylor\_Cordes\_1992}}
\newcommand{\drimmelnir}{\texttt{Drimmel\_NIR\_2000}}
\newcommand{\levine}{\texttt{Levine\_2006}}
\newcommand{\houhan}{\texttt{Hou\_Han\_2014}}
\newcommand{\reid}{\texttt{Reid\_2019}}
\newcommand{\poggio}{\texttt{Poggio\_2021}}
\newcommand{\gaiacollab}{\texttt{Gaia\_2022}}
\newcommand{\drimmelceph}{\texttt{Drimmel\_Ceph\_2024}}

\newcommand{\vallee}{\texttt{Vallee\_1995}}

\newcommand{\python}{\texttt{Python}}
\newcommand{\spiralmap}{\texttt{SpiralMap}}

\begin{document}

\title{SpiralMap: A Python library of the Milky Way's spiral arms}

\author[orcid=0009-0009-6412-4460,sname='Prusty']{Abhay Kumar Prusty}
\affiliation{Indian Institute of Science Education and Research Kolkata, Mohanpur 741246, West Bengal, India}
\email[show]{akp22ms087@iiserkol.ac.in}  

\author[orcid=0000-0000-0000-0002,gname=Shourya, sname='Khanna']{Shourya Khanna} 
\affiliation{INAF - Osservatorio Astrofisico di Torino, via Osservatorio 20, 10025 Pino Torinese (TO), Italy}
\email[]{shourya.khanna@inaf.it}


\section*{Statement of need} 
Mapping the structure of the Galaxy has been an ongoing endeavour for several decades, thanks to which we have come to piece together the components it is made up of such as the discs/bulge/halo and so on \citep{jbhreview2016}. Of particular interest is also the signature of non-axisymmetry in the disc, principally present in the form of the central Galactic bar, the warp, and various spiral like features fanning across a large portion of the disc. Over the years, various groups have surveyed the Galaxy across wavelengths (Radio to Optical) and have deduced the rich variety of spiral structure present in the disc. While some of the papers in literature provide machine readable data to trace out the spiral arms in their model, often this can be a cumbersome exercise for a user simply interested in extracting the coordinates and/or overplotting the spiral arms on another plot of interest, such as while comparing the locations of the arms to features in the velocity field \citep[][etc.]{Khanna:2023,Poggio:2024}. We show an example science application in \autoref{fig:example_science}.\\

With \spiralmap{} we present a library of the major spiral arm models (and maps) of the Galaxy. The package is written in \python{}, and allows the user to both extract the 2D trace and overplot the spiral arms in cartesian/polar coordinates in both Heliocentric (HC) and Galactocentric (GC) frames. A summary of the models currently included is provided in \autoref{tab:description}, where we have tried to include\footnote{Fair practice: If you make use of our package, please ALSO cite the individual model papers. A bib file is provided in \href{https://spiralmap.readthedocs.io/en/latest/citation.html}{\texttt{readthedocs}}.} models from across the electromagnetic spectrum, and based on various tracers (gas/stars etc.). \textbf{Other models can easily be included upon request}. In the near future, we anticipate the availability of 3D spiral arm traces for the Galaxy in literature which can also be included in \spiralmap{}. 
A few example plots that can be generated using the package are included below. For full details, we point to the following links:
a) Documentation: \href{https://spiralmap.readthedocs.io/en/latest/}{\texttt{readthedocs}}, b) Demonstration: \href{https://github.com/Abhaypru/SpiralMap/blob/main/demo_spiralmap.ipynb}{\texttt{Jupyter} notebook}, c) \href{https://github.com/Abhaypru/SpiralMap}{\texttt{GitHub repository}}. 

\begin{deluxetable*}{ll}
\digitalasset
\tablewidth{0pt}
\tablecaption{Basic summary of the spiral arm models included in \spiralmap.\label{tab:description}}
\tablehead{
\colhead{Model} &  \colhead{Description} 
}
\startdata
\taylorcordes & Model based on HII \citep{Taylor:1993}. \\ 
\vallee & Model based on Galactic magnetic field/Dust/stars \citep{Vallee:1995}. \\
\drimmelnir & Model based on Galactic plane emission in the NIR \citep{drimmel2000}. \\ 
\levine & Model based on HI (21 cm) \citep{Levine:2006}. \\
\houhan & Logarithmic spiral model based on HII/ GMC/methanol Maser observations \citep{Hou:2014}. \\ 
\reid & Model based on high-precision parallax measurements of MASERS \citep{Reid:2019}.  \\ 
\poggio & Map based on Upper Main sequence stars \citep{Poggio:2021} \\ 
\gaiacollab & Map based on OB stars \citep{gaiacollab22} \\  
\drimmelceph{} & Model based on Cepheid variables \citep{Drimmel:2024}. \\      
\enddata
\end{deluxetable*}

\clearpage
\begin{figure*}
\plotone{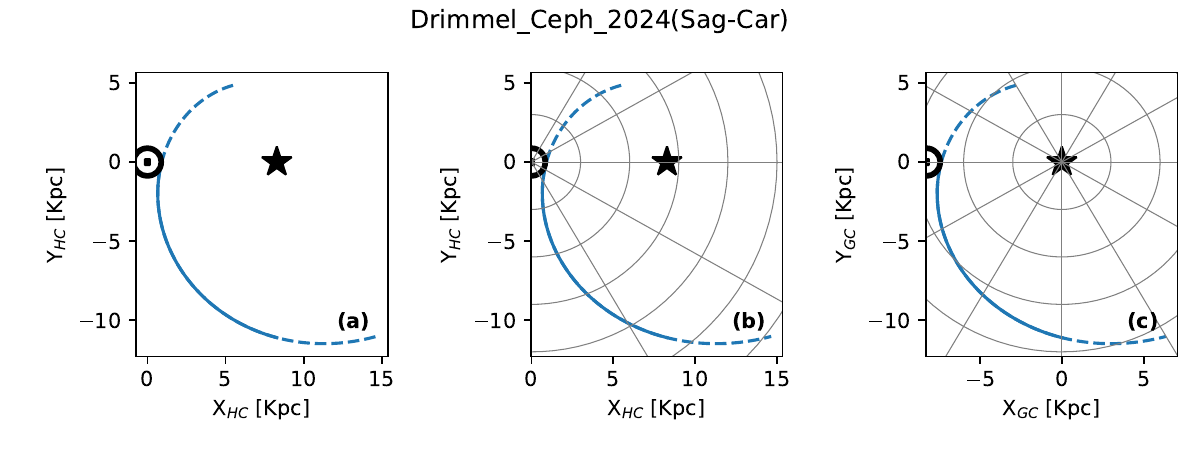}
\caption{Cartesian projection of the \drimmelceph{} model shown for a particular arm (\texttt{Sag-Car}). We show this arm in HC (a), HC with a polar grid in the background (b), and in GC frame with a polar grid in the background (c).}\label{fig:single_arm_single_model}
\end{figure*}

\begin{figure}
\plotone{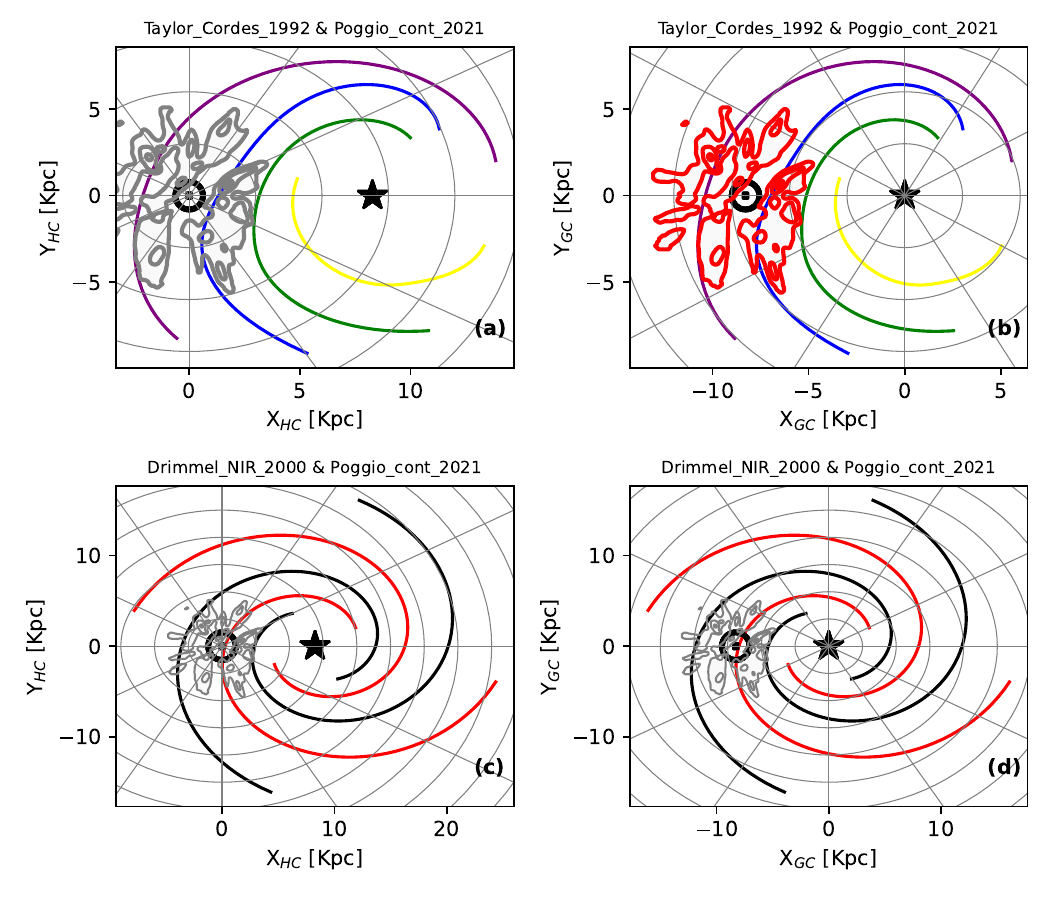}
\caption{Cartesian projections of  multiple models plotted together with a polar grid in the background. We show the \taylorcordes{} \& \poggio{} models in HC (a) and GC (b) frames, and similarly, the \drimmelnir{} \& \poggio{} models in HC (c) and GC (d) frames.}\label{fig:multiple_models_cartesian}
\end{figure}

\begin{figure}
\plotone{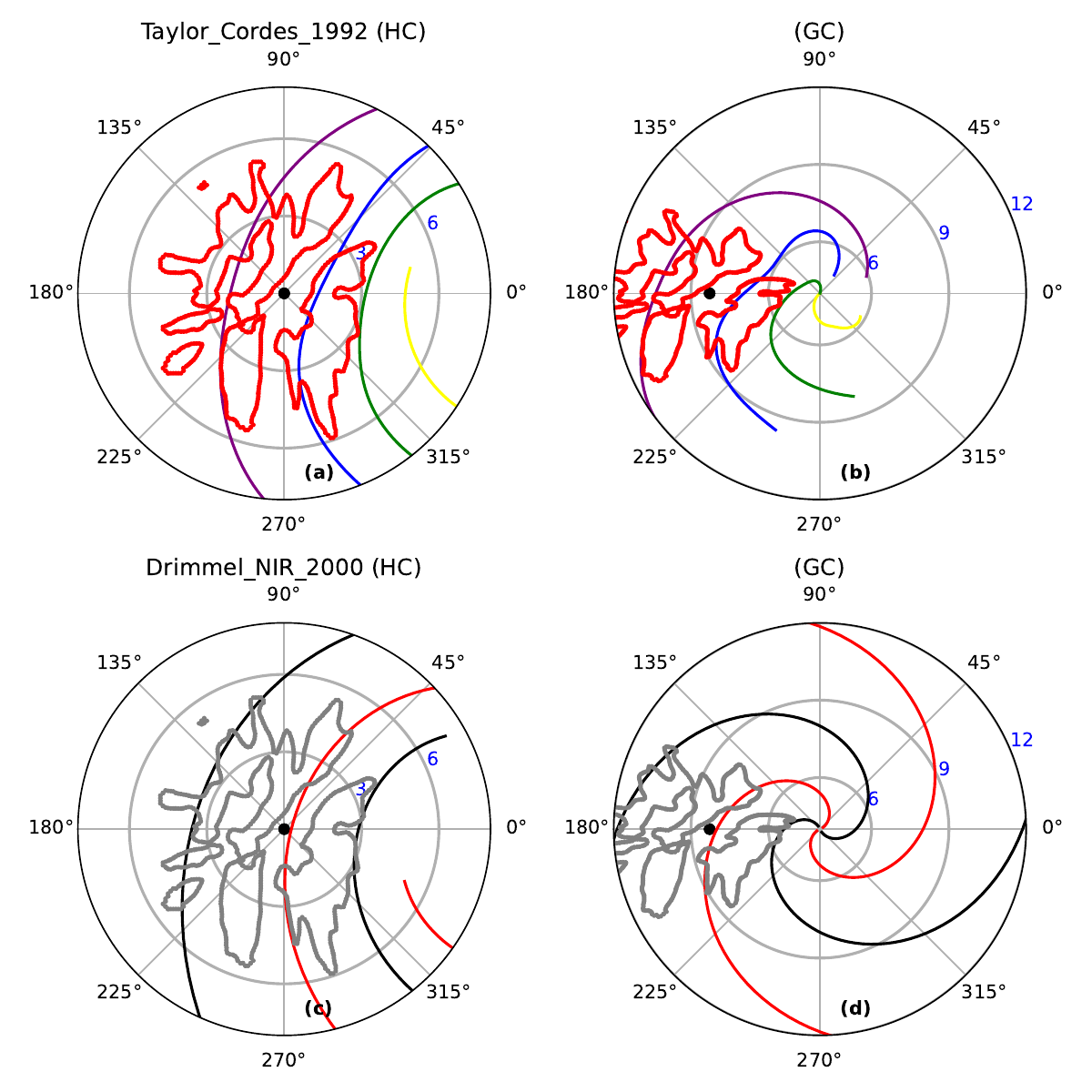}
\caption{Polar projections of multiple models plotted together. We show the \taylorcordes{} \& \poggio{} models in HC (a) and GC (b) frames, and similarly the \drimmelnir{} \& \poggio{} models in HC (c) and GC (d) frames.}
\label{fig:multiple_models_polar}
\end{figure}
\section{Available models}
In the current version, we have included the 9 models described below and summarised in \autoref{tab:description}. The arms in each model (except for contour maps) are given names for convenience.

\begin{itemize}
    \item \textbf{\taylorcordes}: This model is based on radio and optical observations of H II regions. We use the model parameters presented in their Table 1 \citep{Taylor:1993}. There are four arms in this model (Arm1, Arm2, Arm3, Arm4).    
    \item \textbf{\vallee}: This model is based on a combination of the Galactic magnetic field, HI, HII, CO, and Dust distribution. We use their best model as depicted in their figure 3 \citep{Vallee:1995}. There are four arms in this model (Sagittarius, Scutum, Norma, Perseus).
    
    \item \textbf{\drimmelnir}: 
	This model is based on Galactic plane emission profiles in the K band using COBE data. The model is publicly available. There are two arms (1\_arm, 2\_arm) and two inter-arm regions (3\_interarm, 4\_interarm) in this model. 
    
    \item \textbf{\levine}: 
	This model implements the logarithmic spiral framework as described in \cite{Levine:2006}. It is based on HI observations. We have adopted the model from their Table 1. There are four arms in this model (Arm1, Arm2, Arm3, Arm4).

    \item \textbf{\houhan}: 
    This model is built upon the polynomial-logarithmic formulation introduced by \cite{Hou:2014} and is based on a combination of H II, giant molecular clouds, and methanol MASER catalogs. Their model is publicly available. There are six arms in this model (Arm1, Arm2, Arm3, Arm4, Arm5, Arm6). Additionally their local arm is included as a regular logarithmic spiral.

    \item \textbf{\reid}: 
	This model is based on high precision radio astrometry of MASERS \citep{Reid:2019}. We have adopted the model from their Table 2. There are seven arms in this model (3-kpc, Norma, Sct-Cen, Sgr-Car, Local, Perseus, Outer).
	
    \item \textbf{\poggio}: 
     Using Gaia EDR3 astrometry of upper-main sequence stars, \cite{Poggio:2021} generated 2D contour maps of spiral arms. Their data is available publicly, and also included in the package with their permission.
    
    \item \textbf{\gaiacollab}: 
    Using OB stars selected using Gaia Astrophysical parameters \& DR3 astrometry, \cite{gaiacollab22} generated 2D contour maps of spiral arms. Their data is available publicly, and also included in the package with their permission.

    \item \textbf{\drimmelceph}:
    Recently, new distances to 3425 classical cepheids were derived by \cite{skowron2024} with a mean distance uncertainty of about 6\%.  
   From this \cite{Drimmel:2024} used a subsample of dynamically young Milky Way Cepheids to construct models for spiral arms. Their model is publicly available but also included in the package as a userfriendly pickle file, with their permission. There are four arms in this model (Scutum, Sag-Car, Orion, Perseus).
	    
\end{itemize}

This list is by no means exhaustive, but is instead a first attempt to compile spiral arms based on a variety of different tracers.

\section{Example Usage \& Scientific application}

\begin{figure}
\gridline{\fig{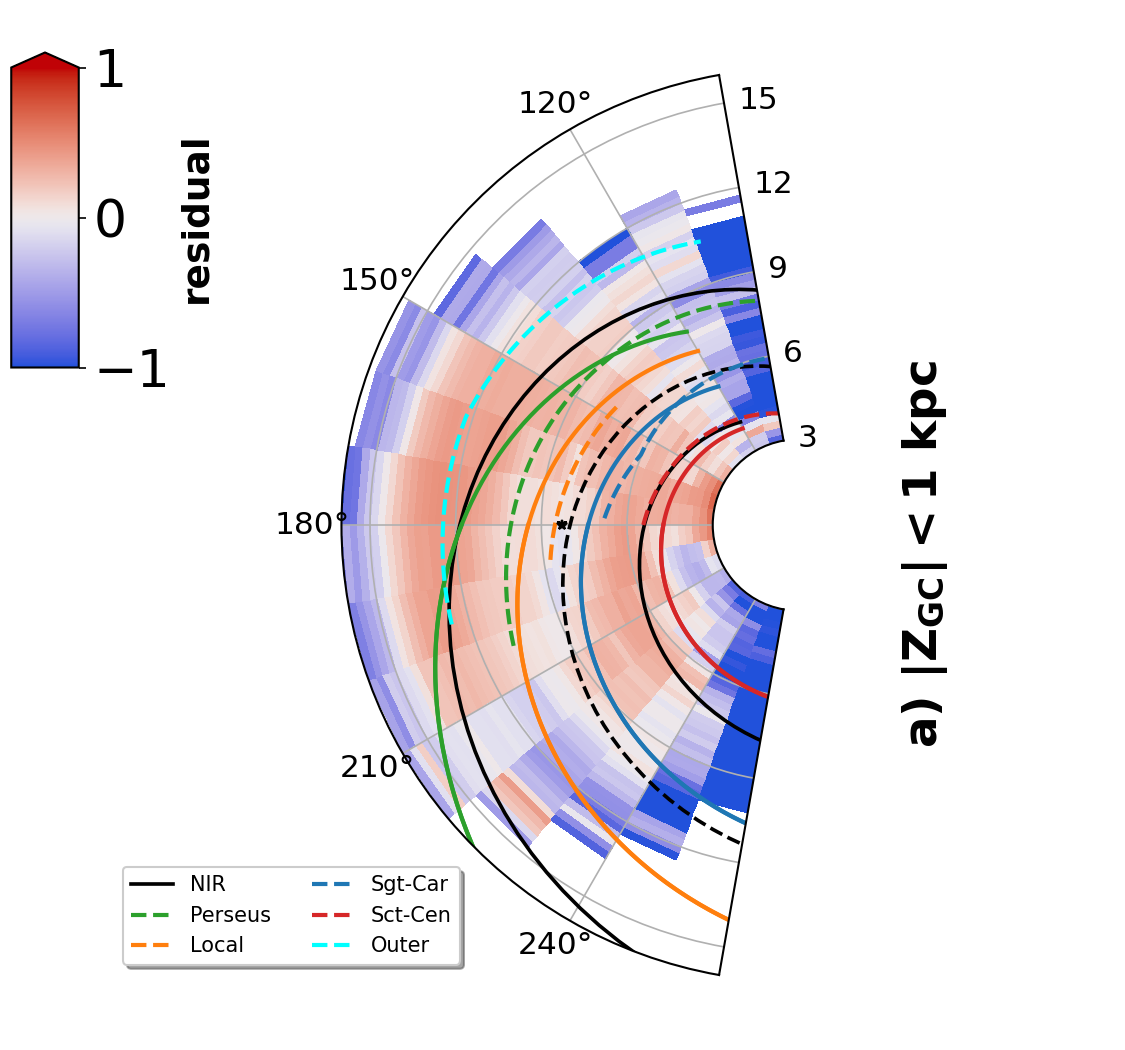}{0.5\textwidth}{(a)}
\fig{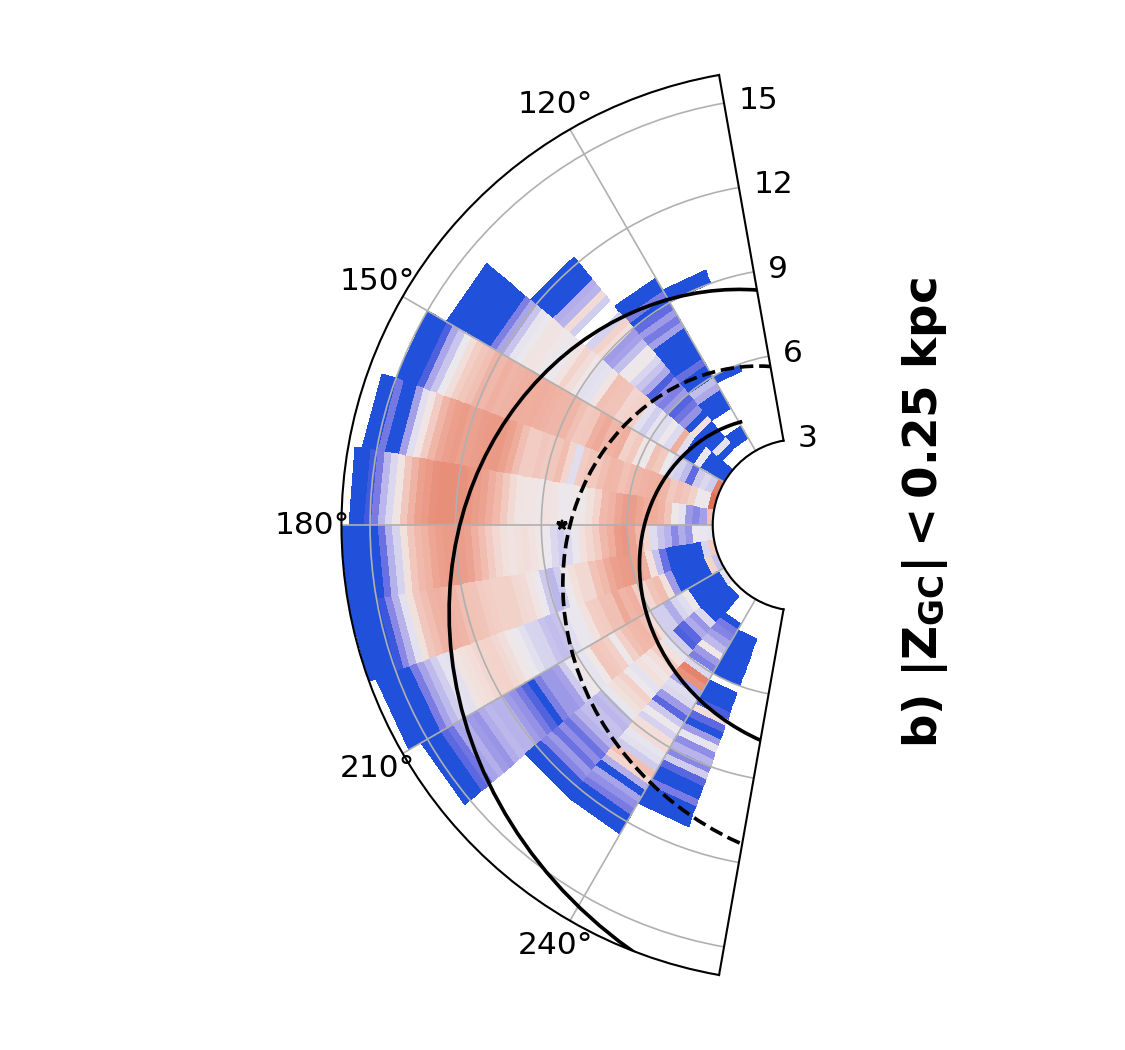}{0.5\textwidth}{(b)}}
\caption{An example science case where \spiralmap{} can be useful. Here we show two figures reproduced from K24 (with permission), comparing the residuals between a best-fit model and data, and the locations of spiral arm models in the Milky Way. Panel(a) shows the \drimmelnir{}, \drimmelceph{}, \& \reid{} overplotted together, while panel(b) shows only the \drimmelnir{} model. Both cases are shown in polar projection.}\label{fig:example_science}
\end{figure}

As an example of a science case, we reproduce figures from \cite[][hereafter K24]{khanna2024} where \spiralmap{} was used. In particular, K24 constructed a model for the stellar density distribution in Red Clump stars in the Milky Way, and then compared the residuals of their best-fit models with the locations of non-axisymmetric structures such as spiral arms. \autoref{fig:example_science} shows two of their residual plots overlaid with spiral models (\drimmelnir{}, \drimmelceph{}, \& \reid{}) using \spiralmap{}.

A very basic example of using \spiralmap{} is shown below, where we access all information about one particular arm ('Sag-Car') in one particular model \drimmelceph,
\begin{verbatim}
##################################################
##### Readout a single arm from a single model ###
##################################################
import SpiralMap as sp
from SpiralMap import polar_style

Rsun=8.277
spirals = sp.main_(Rsun=Rsun)	
use_model = `Drimmel_Ceph_2024'
spirals.getinfo(model=use_model)
plotattrs = {`plot':False}
spirals.readout(plotattrs,model=use_model,arm=`Sag-Car')    
\end{verbatim}

More common examples are shown in the accompanying \href{https://github.com/Abhaypru/SpiralMap/blob/main/demo_spiralmap.ipynb}{\texttt{Jupyter} notebook}. 

\newpage
\begin{acknowledgments}
SK acknowledges support from the European Union's Horizon 2020 research and innovation program under the GaiaUnlimited project (grant agreement No 101004110). We thank Ronald Drimmel \& Eloisa Poggio for useful suggestions.
This work has used the following additional software products:
Matplotlib \citep{Hunter:2007};
IPython \citep{PER-GRA:2007};  
Pandas \citep{reback2020pandas}; 
Astropy, a community-developed core Python package for Astronomy \citep{AstropyCollaboration:2018}; NumPy \citep{harris2020array}.
\end{acknowledgments}


\bibliography{sample7}{}
\bibliographystyle{aasjournal}



\end{document}